\documentstyle[12pt,epsfig,multirow]{article}

\newcommand{\be}{\begin{eqnarray}}
\newcommand{\ee}{\end{eqnarray}}
\newcommand{\etal}{{\it et al.}}
\def\nue{{\nu_e}}
\def\anue{{\bar\nu_e}}
\def\numu{{\nu_{\mu}}}
\def\anumu{{\bar\nu_{\mu}}}
\def\nutau{{\nu_{\tau}}}
\def\anutau{{\bar\nu_{\tau}}}

\newcommand{\ms}{\Delta m^2_{21}}
\newcommand{\ma}{\Delta m^2_{31}}

\newcommand{\sss}{\sin^2 \theta_{12}}
\newcommand{\sch}{\sin^2 \theta_{13}}

\newcommand{\sa}{\sin^2 \theta_{23}}

\newcommand{\sts}{\sin^2 \theta_{34}}

\newcommand{\tbm}{tribimaximal}
\def\ltap{\ \raisebox{-.4ex}{\rlap{$\sim$}} \raisebox{.4ex}{$<$}\ }

\begin{document}

\begin{flushright}
\texttt{HRI-P-07-06-001}\\
\end{flushright}
\bigskip

\begin{center}
{\Large \bf Confusing Sterile Neutrinos with
Deviation from Tribimaximal Mixing at Neutrino Telescopes}

\vspace{.5in}

{\bf Ram Lal Awasthi$^{a}$ and  
Sandhya Choubey$^{b}$}
\vskip .5cm
{\normalsize \it Harish-Chandra Research Institute,} \\
{\normalsize \it Chhatnag Road, Jhunsi, Allahabad  211019, India}
\vskip 2cm

{\bf ABSTRACT}
\end{center}
We expound the impact of extra sterile species on the 
ultra high energy neutrino fluxes in neutrino telescopes.
We use three types of well-known flux ratios and compare the values 
of these flux ratios in presence of sterile neutrinos, with 
those predicted by deviation from the tribimaximal mixing 
scheme. We show that in the upcoming neutrino telescopes, 
its easy to confuse between the signature 
of sterile neutrinos  
with that of the deviation from 
tribimaximal mixing. We also show that 
if the measured flux ratios acquire a
value well outside the range predicted by the standard scenario 
with three active neutrinos only, it might be possible to
tell the presence of extra sterile neutrinos by observing 
ultra high energy neutrinos in future neutrino telescopes.

\vskip 5cm

\noindent $^a$ email: ramlal@mri.ernet.in 

\noindent $^b$ email: sandhya@mri.ernet.in

\newpage

\section{Introduction}

Cumulative effort  
through a series of experiments spanning more than four decades, 
has helped augment
our understanding of the properties of neutrinos. 
That neutrinos have mass, mixing and hence flavor oscillations,
has been firmly established. 
The picture emerging from the combined results of 
solar \cite{solar},
atmospheric \cite{atm}, reactor \cite{kl,chooz},  
and accelerator \cite{k2k,minos} neutrino experiments,
is seen to be consistent with $\ms= 8\times 10^{-5}$ eV$^2$ 
and $\ma=2.5\times 10^{-3}$ 
eV$^2$,
\footnote{We follow the convention where $\Delta m_{ij}^2 =
m_i^2 - m_j^2$.}
and the so-called 
``tribimaximal mixing'' \cite{hps} pattern for the neutrinos
\cite{limits,limits2}. 
In the tribimaximal (TBM) mixing scheme the mixing matrix has the 
form
\be
U^{TBM} = \pmatrix
{\sqrt{\frac{2}{3}} & \sqrt{\frac{1}{3}} & 0\cr
-\sqrt{\frac{1}{6}} & \sqrt{\frac{1}{3}} & \sqrt{\frac{1}{2}}\cr
\sqrt{\frac{1}{6}} & -\sqrt{\frac{1}{3}} & \sqrt{\frac{1}{2}}\cr
} 
~.
\label{eq:upmnstbm} 
\ee
In this scenario, the atmospheric neutrino 
mixing angle $\theta_{23}$
is maximal ($\sa=0.5$), the so-called Chooz mixing 
angle $\theta_{13}=0$, while the solar neutrino mixing 
angle is such that $\sss=1/3$, all consistent with the 
current global data set. 
Maximal $\theta_{23}$ and 
zero $\theta_{13}$ can be easily obtained 
by imposing 
discrete family symmetries such as the $\mu-\tau$ symmetry
\cite{mutau} or the $L_\mu-L_\tau$ symmetry \cite{lmultau}.
While $\mu-\tau$ symmetry does not predict the 
neutrino masses and the solar mixing angle, a 
phenomenologically viable scenario which satisfies 
all neutrino data can be obtained within an approximate 
$L_\mu-L_\tau$ scheme \cite{lmultau}. 

The TBM mixing 
scheme will be put to stringent test with the next generation 
experiments involving neutrinos from the sun,
atmosphere, reactors and 
accelerators. The mixing angle $\sss$ can be 
measured to less than 16\% accuracy at $3\sigma$ using solar neutrinos 
and to 6\% accuracy from a future $\sim 60$ km baseline 
SPMIN (Survival Probability MINimum) reactor experiment 
\cite{th12}. Another prospect for accurate determination 
of $\sss$ is to dope the Super-Kamiokande or even megaton 
water detectors with gadolinium \cite{skgd}.
The small and hitherto undetermined mixing angle $\theta_{13}$ 
will be probed in the up-coming reactor experiments \cite{white},
as well as in 
accelerator based experiments using beams from 
conventional sources
\cite{t2knova}, beams 
produced by decay of accelerated radioactive ions 
stored in rings (``Beta-Beams'') or 
beam from decay of accelerated muons stored in rings 
(``Neutrino Factory'') \cite{nufact}. In principle, one of 
the best determination of $\theta_{13}$ could be possible
from the neutrino signal of a future galactic supernova
\cite{sn}.  
The 
deviation of $\theta_{23}$ from maximality and the sign of 
$D_{23} \equiv 0.5 -\sa$ 
can be experimentally checked in 
atmospheric neutrino experiments \cite{th23}.

A very well known and interesting feature of the TBM
mixing scheme arises in the context of the flavor ratios 
of ultra high energy neutrinos arriving on earth. 
Ultra high energy neutrinos are created mainly through 
decay of high energy pions produced in $pp$ and $p\gamma$ 
collisions. Hence their relative flavor content at the source is 
expected to be
\be
\bigg\{\phi_e^0:\phi_\mu^0:\phi_\tau^0\bigg\} \equiv \bigg\{1:2:0\bigg\}~,
\ee
where $\phi_\alpha^0$ ($\alpha\equiv e$, $\mu$, $\tau$) 
are the fluxes at the source. 
Under the TBM scheme, due to the inherent 
$\mu-\tau$ symmetry, $\theta_{23}=45^\circ$ and $\theta_{13}=0$, 
and therefore one obtains the flux ratio at earth as
\cite{flavor}
\be
\bigg\{\phi_e:\phi_\mu:\phi_\tau\bigg\} \equiv \bigg\{1:1:1\bigg\}~.
\label{eq:flavorearth}
\ee

Next generation km$^3$ neutrino telescopes such as the IceCube 
in Antarctica \cite{icecube}, Km3NET in the 
Mediterranean \cite{km3net} 
are being especially designed and built to detect ultra high 
energy neutrinos coming from astrophysical sources. 
A lot of interest has been recently generated on the 
potential of using the 
observed flavor ratios of the ultra high energy neutrinos 
in neutrino telescopes in deciphering the  
predictions or deviation from \tbm{} mixing 
\cite{tbmnutel}.  The key idea is that the flavor ratios 
are predicted to be one, as given by Eq. (\ref{eq:flavorearth}),
only if $\theta_{13}=0$ and $\theta_{23}=45^\circ$, a prediction 
of TBM mixing. If the mixing matrix was to deviate slightly 
from the TBM prediction through a change in $\theta_{23}$ or 
$\theta_{13}$, we would see a difference in the flavor 
ratios, which could be used to pin down the extent of this 
deviation. 

Here we study how the flavor ratios 
of ultra high energy neutrinos get affected if 
we have extra sterile neutrinos mixed with the three active ones.
Sterile neutrinos have been the subject of much discussion recently,
following the much-awaited MiniBooNE results \cite{miniboone}. 
The MiniBooNE experiment was designed to specifically test the 
observed signal at the LSND experiment \cite{lsnd}, 
which can be explained 
most convincingly by neutrino oscillations. The 
MiniBooNE data seem to contradict the LSND signal, as they 
do not see the kind of electron excess predicted by 
the oscillation explanation of the LSND data sample. 
Since the claimed flavor oscillation observed by LSND demands a 
$\Delta m^2 \sim $ eV$^2$, it cannot be accommodated along with the 
solar and atmospheric neutrino observations within a three-generation 
framework. Three separate $\Delta m^2$ scales can be possible 
if we have at least 4 neutrino states. Since the number of light 
active neutrinos are restricted to three from the decay width of 
the $Z$ bosons at LEP, the extra neutrinos have 
to be sterile. The most economical scenario, with only one extra 
sterile neutrino leads to the so-called 2+2 and 3+1 mass 
spectra for the neutrinos \cite{sterileold}. 
The 2+2 spectrum is heavily disfavored by the solar and atmospheric 
neutrino data \cite{limits2}. The 3+1 mass spectrum 
though comparatively less disfavored before the 
MiniBooNE results, 
suffers from a tension in explaining 
simultaneously the observed 
oscillation signal in the LSND experiment and the 
null signals in the other short baseline experiments \cite{sbl}.
The full analysis of all short baseline data including the 
MiniBooNE results further disfavors the 3+1 scheme 
\cite{maltonischwetz}. Adding two sterile neutrinos 
turns out to be the next option to reconcile the LSND 
observations with the rest of the world neutrino data.
In this so-called 3+2 scheme \cite{threeplustwo} 
the tension between the LSND and MiniBooNE data is reduced 
if one allows for CP violation, and this 
mass spectra is found to be consistent with global neutrino 
oscillation results \cite{maltonischwetz}.  

In this paper we look at the signatures of extra sterile neutrinos 
on the observed flavor ratios of ultra high energy neutrinos 
in neutrino telescopes. We look at the impact of sterile neutrinos 
on the flavor ratios {\it vis-a-vis} the impact due to deviation 
from TBM mixing. 
Firstly, for small active-sterile mixing, 
we elucidate the confusion between the 
two cases and 
identify the range of the mixing angles between the 
active and sterile neutrinos which would give flavor ratios 
close to the ones predicted by deviated 
TBM mixing scenarios.
Next we show that large values of hitherto unconstrained mixing 
angles between the active and sterile neutrino species 
could lead to extreme values for flavor ratios, providing 
smoking gun signal for the existence of light sterile 
neutrinos which are heavily mixed with the active neutrinos.
Thus signal from neutrino telescopes could provide an 
independent check of the results from the short baseline 
experiments in general and LSND and MiniBooNE in particular.

We present all results in this paper in the framework of 3+1 scenario 
{\it just for simplicity}. It is absolutely
straightforward to extend our results and conclusions to the 
3+2 scenario. Here we have chosen not to work with the 3+2 mass spectra 
because that would 
entail many extra mixing angles, making the results look 
complicated. Therefore for the sake of illustration, we allow only 
two active sterile mixing angles to be non-zero in this work. 
In section 2
we give a brief discussion of the ultra high energy neutrino 
fluxes and the corresponding flux/flavor ratios. 
In section 3 the prediction from TBM mixing and the 
impact of deviation from TBM mixing in the framework of 
just three active neutrinos is reviewed. 
Section 4 gives the oscillation probabilities for 
three active and one extra sterile neutrino, while 
section 5 gives the corresponding flux ratios. In section 6
we present our numerical results. We end in section 7 with 
our conclusions.

\section{Ultra high energy neutrino fluxes}

Ultra high energy neutrinos
are predicted from a number of 
astrophysical sources. The {\it fireball} model which has been
put forward as the most plausible source for 
gamma ray bursts, 
are expected to produce ultra high energy neutrinos 
as well. Protons moving in the jets of the {\it fireball} 
are accelerated to very high energies. These highly 
accelerated protons undergo $pp$ (or $pn$) and $p\gamma$ 
collisions, producing pions. The pions produce neutrinos through 
their decay channel
\be
\pi^{+} \longrightarrow \mu^{+} + \nu_{\mu}  ~, 
~~{\rm followed~by}~ ~~ 
\mu^{+}\longrightarrow e^{+} + \nu_{e} + \anumu  
\ee
and
\be
\pi^{-} \longrightarrow \mu^{-} + \anumu~,
~~{\rm followed~by}~ ~~
\mu^{-} \longrightarrow e^{-} + \anue + \nu_{\mu}~.
\ee
It is expected that 
the $p\gamma$ reaction will predominantly produce $\pi^+$, 
while both $\pi^+$ and $\pi^-$ are expected from the $pp$ and 
$pn$ collisions. Therefore, for the $pp$ and $pn$ collisions we 
expect the flavor ratio 
\be
\bigg\{\phi_\nue^0:\phi_{\anue}^0:\phi_{\numu}^0:\phi_{{\anumu}}^0:
\phi_\nutau^0:\phi_{{\anutau}}^0:\bigg\} \equiv 
\bigg\{1:1:2:2:0:0\bigg\}~,
\ee
where $\phi_{\nu_\alpha}^0$ and $\phi_{\bar\nu_{\alpha}}^0$ are the 
neutrino and antineutrino fluxes respectively, 
of flavor $\alpha$. For $p\gamma$ collisions on the other hand,
we expect
\be
\bigg\{\phi_\nue^0:\phi_{\anue}^0:\phi_{\numu}^0:\phi_{{\anumu}}^0:
\phi_\nutau^0:\phi_{{\anutau}}^0:\bigg\} \equiv \bigg\{1:0:1:1:0:0\bigg\}~.
\ee
Though apparently it looks that the flux ratio for the two 
dominant channels of ultra high energy neutrinos are different, 
the neutrino telescopes will not have 
sensitivity to the charge of the resultant leptons and hence 
in general are not expected to be able to distinguish the 
neutrino of a given flavor from its antineutrino. Only when 
the Glashow resonance ($\anue e \rightarrow W^- \rightarrow $ anything) 
is the detection channel, 
$\anue$ is unambiguously detected and it is possible to measure 
the $\nue$ and $\anue$ signal separately. However, this occurs 
only over a small energy window $E = m_W^2/2m_e=6.3$ PeV, 
where $m_W$ and $m_e$ are the $W$ and $e$ mass respectively.
In what follows, we 
will consider the sum of the neutrino and antineutrino signal 
at the neutrino telescope and 
in general refer to the flavor ratio as 
\be
\bigg\{\phi_e^0:\phi_\mu^0:\phi_\tau^0\bigg\} \equiv \bigg\{1:2:0\bigg\}~,
\label{eq:flux}
\ee
where $\phi_\alpha \equiv \phi_{\nu_\alpha} + \phi_{\bar \nu_\alpha}$.
If the ultra high energy neutrinos come from a neutron source, 
then the flux ratio is expected to be
\be
\bigg\{\phi_\nue^0:\phi_{\anue}^0:\phi_{\numu}^0:\phi_{{\anumu}}^0:
\phi_\nutau^0:\phi_{{\anutau}}^0:\bigg\} \equiv \bigg\{0:1:0:0:0:0\bigg\}~,
\ee
while for a muon damped source where the secondary muon is trapped 
and hence does not decay, it would be
\be
\bigg\{\phi_\nue^0:\phi_{\anue}^0:\phi_{\numu}^0:\phi_{{\anumu}}^0:
\phi_\nutau^0:\phi_{{\anutau}}^0:\bigg\} \equiv \bigg\{0:0:1:1:0:0\bigg\}~.
\ee
However, in this paper we will assume that only pions are the sources of 
ultra high energy neutrinos and work with the flavor ratio given 
in Eq. (\ref{eq:flux}).

Since the  absolute flux predictions for the ultra high energy 
neutrinos could be uncertain by a huge amount, it is better to work 
with ratios of the fluxes. 
In this paper we exemplify the predictions of 
deviation from TBM mixing, and the impact of an additional 
sterile neutrino, using the following flux ratios:
\be
R_e = \frac{\phi_e}{\phi_\mu+\phi_\tau}~,
~~~
R_\mu = \frac{\phi_\mu}{\phi_e+\phi_\tau}~,
~~~
R_\tau = \frac{\phi_\tau}{\phi_e+\phi_\mu}~,
\ee
where $\phi_e$, $\phi_\mu$ and $\phi_\tau$ 
respectively are the 
$\nue+\anue$, $\numu+\anumu$ and $\nutau+\anutau$ 
fluxes at earth after oscillations. Written in terms 
of the oscillation probabilities:
\be
R_e = \frac{P_{ee}+2\,P_{e\mu}}{2P_{\mu\mu}
+P_{e\mu}+P_{e\tau} + 2P_{\mu\tau}},
~
R_\mu =  \frac{2\,P_{\mu\mu}+P_{e\mu}}{P_{ee}
+2P_{e\mu}+P_{e\tau} + 2P_{\mu\tau}},
~
R_\tau =  \frac{2\,P_{\mu\tau}+P_{e\tau}}{P_{ee}
+2P_{e\mu}+2P_{\mu\mu} + P_{e\mu}},
\label{eq:r}
\ee
where we have used $P_{e\mu}=P_{\mu e}$, since there is 
no chance of CP violation here. 
For the case for three active neutrinos only the 
flux ratios are given by
\be
R_e = \frac{1+(P_{e\mu}-P_{e\tau})}
{2-(P_{e\mu}-P_{e\tau})}~,
~~~
R_\mu = \frac{2-(P_{e\mu}+2\,P_{\mu\tau})}
{1+(P_{e\mu}+2\,P_{\mu\tau})}~,
~~~
R_\tau = \frac{P_{e\tau}+2\,P_{\mu\tau}}
{3-(P_{e\tau}+2\,P_{\mu\tau})}~,
\label{eq:ractive}
\ee
where $P_{\alpha\beta}$ is the $\nu_\alpha \rightarrow 
\nu_\beta$ oscillation probability discussed in the 
following sections. For the case where we have one extra 
sterile neutrino in addition to the three active ones,
the flux ratios can be written as
\be
R_e = \frac{1+(P_{e\mu}-P_{e\tau})-P_{es}}
{2-(P_{e\mu}-P_{e\tau})-2P_{\mu s}},
R_\mu = \frac{2-(P_{e\mu}+2P_{\mu\tau})-2P_{\mu s}}
{1+(P_{e\mu}+2P_{\mu\tau})-P_{es}},
R_\tau = \frac{P_{e\tau}+2\,P_{\mu\tau}}
{3-(P_{e\tau}+2P_{\mu\tau})-P_{es}-2P_{\mu s}},
\nonumber
\label{eq:rsterile}
\ee
where $P_{es}$ and $P_{\mu s}$ are the $\nue\rightarrow \nu_s$
and $\numu\rightarrow \nu_s$ oscillation probabilities 
respectively. 
However, we prefer to express the flux 
ratios in the sterile case by the most general form
given by Eq. (\ref{eq:r}).

\section{Oscillation probabilities and flux ratios in TBM mixing}

For neutrinos coming from astrophysical sources the 
oscillation probabilities can be written in general as
\be
P_{\alpha\beta} &=& \delta_{\alpha\beta} - \,\sum_{i\neq j}
U_{\alpha i}U_{\beta i}^*U_{\alpha j }^*U_{\beta j}
\nonumber\\
&=& \sum_{i}|U_{\alpha i}|^2\,|U_{\beta i}|^2
~,
\label{eq:probgeneral}
\ee
where $i,j$ and $\alpha,\beta$
are the generation indices of the mass basis and 
flavor basis respectively. 
The oscillatory terms, $\sin^2(\Delta m^2_{ij}L/4E)$, have 
averaged out to 1/2 and the hence the probabilities 
depend only on the elements of the mixing matrix. 
The neutrino mixing matrix for three active neutrinos
can be parameterized as \cite{PMNS}
\be
U = \pmatrix
{C_{12} C_{13} & S_{12} C_{13} & S_{13} e^{-i \delta} 
\cr
-S_{12} C_{23} - C_{12} S_{23} S_{13} e^{i \delta} 
& C_{12} C_{23} - S_{12} S_{23} S_{13} e^{i \delta} 
& S_{23} C_{13} \cr
S_{12} S_{23} - C_{12} C_{23} S_{13} e^{i \delta} 
& 
-C_{12} S_{23} - S_{12} C_{23} S_{13} e^{i \delta} 
& C_{23} C_{13}\cr
} 
~,
\label{eq:upmns} 
\ee 
where $C_{ij}=\cos\theta_{ij}$ and $S_{ij}=\sin\theta_{ij}$,
and $\delta$ the CP phase.
Under the \tbm{} mixing ansatz, $S_{12}^2=1/3$, $S_{23}^2=1/2$ and
$S_{13}^2=0$, hence 
Eq. (\ref{eq:upmns}) reduces to the 
form given by Eq. (\ref{eq:upmnstbm}). 
The oscillation probabilities therefore have an extremely 
simple form and are given as
\be
P_{ee}^{TBM} = \frac{5}{9},
~~P_{\mu\mu}^{TBM} &=&\frac{7}{18},
~~P_{e\mu}^{TBM} =\frac{2}{9},
~~P_{e\tau}^{TBM} =\frac{2}{9},
~~P_{\mu\tau}^{TBM} =\frac{7}{18},
~~P_{\tau\tau}^{TBM} = \frac{7}{18}.
\label{eq:prtbm}
\ee
The effect of the $\mu-\tau$ symmetry in-built in TBM 
mixing is
clearly reflected by $P_{\mu\mu}^{TBM}=P_{\mu\tau}^{TBM}=P_{\tau\tau}^{TBM}$
and   
$P_{e\mu}^{TBM}=P_{e\tau}^{TBM}$ 
in Eq. (\ref{eq:prtbm}). 
Therefore, the flux ratios 
defined in the previous section are 
\be
R_e^{TBM} = R_\mu^{TBM} = R_\tau^{TBM} = \frac{1}{2} ~.
\ee
Any deviation from TBM mixing 
in the three active neutrino case will be 
reflected in these flux ratios, which 
will deviate from 1/2. We calculate the probabilities using 
Eqs. (\ref{eq:probgeneral}) and (\ref{eq:upmns}) for the
case where mixing angles are different from those predicted 
by TBM mixing and present the results for the flux ratios
using Eq. (\ref{eq:ractive}).


\section{Oscillation probabilities with sterile neutrinos}

In what follows, 
we will work in a framework where we include one extra 
sterile neutrino in addition to the three active ones and 
parameterize the $4\times 4$ mixing matrix as
\be
U_s = R(\theta_{34})R(\theta_{24})R(\theta_{23})
R(\theta_{14})R(\theta_{13})R(\theta_{12})~,
\label{eq:uspara}
\ee
where $R(\theta_{ij})$ are the rotation matrices 
and $\theta_{ij}$ the mixing angle. 
Note that we have put all phases to zero in Eq. (\ref{eq:uspara}).
In general for the 3+1 scenario there are 3 CP violating 
Dirac phases.
However, in this paper we 
have neglected the effect of the 
CP violating phases for simplicity.

\subsection{TBM mixing with sterile neutrinos}

If we assume that the mixing angles $\theta_{12}$,
$\theta_{13}$ and $\theta_{23}$ follow the same values 
as in TBM mixing, {\it i.e.}, $\sss = 1/3$, $\sch=0$ and 
$\sa=1/2$,  and we further assume that the 
mixing angle $\theta_{14}=0$, 
then the  $4\times 4$ mixing matrix $U_s$ is given 
by
\be
U_s = 
\bordermatrix{& & &   &\cr
          & \sqrt{\frac{2}{3}}  & \sqrt{\frac{1}{3}}  & 0  & 0  \cr
           & -\sqrt{\frac{1}{6}}C_{24}  & \sqrt{\frac{1}{3}}C_{24}  
& \sqrt{\frac{1}{2}}C_{24}  & S_{24}  \cr
           & \sqrt{\frac{1}{6}}[C_{34}+S_{24}S_{34}] 
& -\sqrt{\frac{1}{3}}[C_{34}+S_{24}S_{34}]  
& \sqrt{\frac{1}{2}}[C_{34}-S_{24}S_{34}]  & C_{24}S_{34}  \cr
           & \sqrt{\frac{1}{6}}[C_{34}S_{24}-S_{34}]  
& -\sqrt{\frac{1}{3}}[C_{34}S_{24}-S_{34}]  
& -\sqrt{\frac{1}{2}}[C_{34}S_{24}+S_{34}]  & C_{24}C_{34}  \cr}
~.
\ee
Using Eq. (\ref{eq:probgeneral}) 
we get the oscillation probabilities:
\be
P_{ee}&=&1-2   U_{e2}  ^{2}   U_{e1}  ^{2}~,
\nonumber\\
&=& \frac{5}{9}~,
\\
\nonumber\\
P_{e\mu}&=&-2 U_{e2} U_{e1} U_{\mu2} U_{\mu1}~,
\nonumber\\
&=& \frac{2}{9}C_{24}^{2}~,
\\
\nonumber\\
P_{\mu\mu}&=&1-2  U_{\mu2}^{2}   U_{\mu1}
   ^{2}-2   U_{\mu3}  ^{2}\left(   U_{\mu2}  ^{2} +  
  U_{\mu1}  ^{2}\right) - 2   U_{\mu4}  ^{2}\left( 1-  U_{\mu4}
   ^{2}\right)~,
\nonumber\\
&=& 1-2 C_{24}^{2}+\frac{25}{18}C_{24}^{4}~,
\\
\nonumber\\
P_{e\tau}&=&-2 U_{e2}U_{e1}U_{\tau2}U_{\tau1}~,
\nonumber\\
&=& \frac{2}{9}\left( C_{34}+S_{24}S_{34}\right)^{2}~,
\\
\nonumber\\
P_{\mu\tau}&=&-2 U_{\mu2}U_{\mu1}U_{\tau2}U_{\tau1}-2
  U_{\mu3}U_{\tau3}\left(U_{\mu2}U_{\tau2}+U_{\mu1}U_{\tau1}\right)+ 
2 U_{\mu4}^{2}   U_{\tau4}  ^{2} ~,
\nonumber\\
&=& \frac{1}{18}\left( 7 C_{24}^{2}C_{34}^{2}+25 
C_{24}^{2}S_{24}^{2}S_{34}^{2}-4 C_{24}^{2}C_{34}S_{24}S_{34}\right)~.
\ee

\subsection{Deviation from TBM mixing and sterile neutrinos}

If the mixing angles $\theta_{12}$, $\theta_{13}$ and $\theta_{23}$ 
are 
different from that 
predicted by the exact TBM ansatz, then one has to use the 
expression for the full $U_s$ given by Eq. (\ref{eq:uspara}). 
For simplicity we use just one case where the mixing in 
the $3\times 3$ active sector 
could deviate from TBM, {\it viz}, the case where the mixing 
angle $\theta_{23}$ is non-maximal. For this case the matrix 
$U_s$ is given by
\be
U_s = 
\bordermatrix{& & &   &\cr
          & \sqrt{\frac{2}{3}}  & \sqrt{\frac{1}{3}}  & 0  & 0  \cr
           & -\sqrt{\frac{1}{3}}C_{23}C_{24}  & \sqrt{\frac{2}{3}}
C_{23}C_{24}  & S_{23}C_{24}  & S_{24}  \cr
           & \sqrt{\frac{1}{3}}[S_{23}C_{34}+C_{23}S_{24}S_{34}] &
           -\sqrt{\frac{2}{3}}[S_{23}C_{34}+C_{23}S_{24}S_{34}]  &
           [C_{23}C_{34}-S_{23}
S_{24}S_{34}]  & C_{24}S_{34}  \cr
           & \sqrt{\frac{1}{3}}[C_{23}C_{34}S_{24}-S_{23}S_{34}]  &
           -\sqrt{\frac{2}{3}}[C_{23}C_{34}S_{24}-S_{23}S_{34}]  &
           -[C_{34}S_{23}S_{24}+C_{23}S_{34}]  & C_{24}C_{34}  \cr}
~,
\ee
and the oscillation probabilities are
\be
P_{ee}&=&5/9~,
\\
P_{e\mu}&=&\frac{4}{9}C_{23}^{2}C_{24}^{2}~,
\\
P_{\mu\mu}&=&1-\frac{4}{9}C_{23}^{4}C_{24}^{4}-2C_{24}^{4}S_{23}^{2}
C_{23}^{2}-2S_{24}^{2}C_{24}^{2}~,
\\
P_{e\tau}&=&\frac{4}{9}\left(C_{34}S_{23}+S_{24}S_{34}C_{23}\right)^{2}~,
\\
P_{\mu\tau}\!\!&=&\!\!\frac{1}{36}C_{24}^{2}\left[ 56 C_{23}^{2} 
C_{34}^{2} S_{23}^{2}+
S_{24}\{(57-8 C2_{23}+7 C4_{23}) S_{24}S_{34}^{2}+2(-2+7
C2_{23})S2_{23}S2_{34}\}\right]~,
\ee
where $C2_{ij}=\cos2\theta_{ij}$ and $S2_{ij}=\sin2\theta_{ij}$.
In this paper we will mainly discuss the deviation from TBM 
mixing by changing $\sa$ from its maximal value.  
The TBM mixing ansatz is also violated if either $\sch \neq 0$ 
or $\sss \neq 1/3$. We will present one plot where we show
the flux ratios as a function 
of $\sch$. We will see that the effect of deviation of the 
flux ratios from the TBM predicted value of 1/2 due to $\sch$ 
is very small compared to the change due to $\sa$. 
We therefore keep 
$\sch$ fixed at zero for the results predicted by the sterile 
case in this paper. 
\footnote{Since $\sch=0$ the effect of the normal CP phase is also
completely absent.} 
Therefore for all results with sterile neutrinos, 
we will keep $\sch=0$ and $\sss=1/3$ for simplicity 
and allow only
$\sa$ to deviate from 
maximality.

\section{Flux ratios with sterile neutrinos}

As discussed before, flux ratios for ultra high energy 
neutrinos are the best model independent probes for 
understanding neutrino properties. 
If we consider the case where the only deviation for TBM mixing
comes from $\theta_{23}$ being non-maximal, while $\sss=1/3$ and 
$\theta_{13}=0$, and continue to work under the approximation 
that $\theta_{14}=0$, then  
the flux ratios $R_e$, $R_\mu$ and $R_\tau$ are given as
\begin{eqnarray}
R_{e}=\frac{5+8C_{24}^{2}C_{23}^{2}}
{R_e^{denom}}~,
\end{eqnarray}
where,
\begin{eqnarray}
R_e^{denom}&=&
(18+4C_{24}^{2}C_{23}^{2}-8C_{23}^{4}C_{24}^{4}-36C_{24}^{4}S_{23}^{2}
C_{23}^2-36S_{24}^{2}C_{24}^{2}+4(C_{34}S_{23}+S_{24}S_{34}C_{23})^{2}
\nonumber\\
&-&
8C_{24}^{2}C_{23}^{2}(C_{34}S_{23}+S_{34}S_{24}C_{23})^{2}+36C_{24}^{2}
S_{23}C_{23}(C_{23}C_{34}-S_{23}S_{24}S_{34})(S_{24}S_{34}C_{23}+S_{23}C_{34})
\nonumber\\
&+&
36S_{24}^{2}S_{34}^{2}C_{24}^{2})~.
\end{eqnarray}

\begin{eqnarray}
R_{\mu}=\frac{18+4C_{24}^{2}C_{23}^{2}-8C_{23}^{4}
C_{24}^{4}-9C_{24}^{4}S2_{23}^{2}-9S2_{24}^{2}}
{R_\mu^{denom}}~,
\end{eqnarray}
where,
\begin{eqnarray}
R_\mu^{denom}&=&
5+8C_{24}^{2}C_{23}^{2}+4(C_{34}S_{23}+S_{24}S_{34}C_{23})^{2}+
9S_{34}^{2}S2_{24}^{2}-8C_{24}^{2}C_{23}^{2}(C_{34}S_{23}+
S_{24}S_{34}C_{23})^{2}
\nonumber\\
&+&
18C_{24}^{2}S2_{23}(C_{34}C_{23}-S_{23}S_{24}S_{34})
(S_{24}S_{34}C_{23}+S_{23}C_{34})~,
\end{eqnarray}
and
\begin{eqnarray}
R_{\tau}=\frac{R_\tau^{num}}
{23+12C_{23}^{2}C_{24}^{2}-8C_{24}^{4}C_{23}^{4}-
9C_{24}^{4}S2_{23}^{2}-9S2_{24}^{2}}~,
\end{eqnarray}
where,
\begin{eqnarray}
R_\tau^{num}&=&
4(C_{34}S_{23}+S_{34}S_{24}C_{23})^{2}+9S_{34}^{2}
S2_{24}^{2}-8C_{24}^{2}C_{23}^{2}(C_{34}S_{23}+S_{34}S_{24}C_{23})^{2}
\nonumber\\
&+&
18C_{24}^{2}S2_{23}(C_{34}C_{23}-S_{23}S_{24}
S_{34})(S_{24}S_{34}C_{23}+S_{23}C_{34})~.
\end{eqnarray}

\subsection{Flux ratios when $\theta_{24}=0$}

If we work in a 
further simplified scenario where the only non-zero 
sterile mixing angle is the angle $\theta_{34}$ and 
put both $\theta_{14}=0$ and $\theta_{24}=0$, then the flux ratios
are given as 
\begin{eqnarray}
R_e &=& \frac{5+8C_{23}^{2}}{22-4S_{23}^{2}S_{34}^{2}-
8C_{23}^{4}-36S_{23}^{2}C_{23}^{2}+28S_{23}^{2}C_{23}^{2}C_{34}^{2}}~,
\label{eq:re_24zero}
\end{eqnarray}
\begin{eqnarray}
R_{\mu}&=& \frac{18+4C_{23}^{2}-8C_{23}^{4}-9S2_{23}^{2}}
{5+8C_{23}^{2}+4C_{34}^{2}S_{23}^{2}+
7S2_{23}^{2}C_{34}^{2}}~,
\label{eq:rmu_24zero}
\end{eqnarray}
\begin{eqnarray}
R_{\tau}&=& \frac{4C_{34}^{2}S_{23}^{2}+28C_{23}^{2}
S_{23}^{2}C_{34}^{2}}
{23-24C_{23}^{2}S_{23}^2+4C_{23}^4}~.
\label{eq:rtau_24zero}
\end{eqnarray}

\subsection{Flux ratios when $\theta_{34}=0$}

If instead of putting $\theta_{24}=0$, we put 
$\theta_{34}=0$ along with $\theta_{14}=0$ then the 
flux ratios are
\begin{eqnarray}
R_{e}=\frac{5+8C_{24}^{2}C_{23}^{2}}{22-4C_{23}^{2}-
36C_{24}^{2}S_{24}^2+32C_{24}^{2}C_{23}^{2}-28C_{24}^{2}S_{24}^2
C_{23}^{4}-36C_{24}^{4}C_{23}^{2}}~,
\end{eqnarray}
\begin{eqnarray}
R_{\mu}&=& \frac{18+4C_{24}^{2}C_{23}^{2}-8C_{23}^{4}
C_{24}^{4}-9C_{24}^{4}S2_{23}^{2}-9S2_{24}^{2}}
{9-28C_{24}^{2}C_{23}^{4}-4C_{23}^{2}+36C_{24}^{2}C_{23}^{2}}~,
\end{eqnarray}
\begin{eqnarray}
R_{\tau}=\frac{4S_{23}^{2}+28C_{24}^{2}C_{23}^{2}S_{23}^2}
{23+12C_{23}^{2}C_{24}^{2}-8C_{24}^{4}C_{23}^{4}-
9C_{24}^{4}S2_{23}^{2}-9S2_{24}^{2}}~.
\end{eqnarray}
Since the constraints from the short baseline experiments restrict 
$\theta_{24}$ to very small values, we can 
expand the flux ratios in a Taylor series and keep only the first 
order terms in $\sin^2\theta_{24}$, giving
\begin{eqnarray}
R_{e}\simeq\frac{5+8C_{24}^{2}C_{23}^{2}}{22-8C_{23}^{2}-36S_{24}^{2}+
12C_{23}^{4}S_{24}^{2}+10S2_{23}^{2}S_{24}^{2}}~,
\label{eq:re24}
\end{eqnarray}
\be
R_{\mu}\simeq \frac{(18-32C_{23}^{2}+28C_{23}^{4})-
4S_{24}^{2}(9-17C_{23}^{2}+14C_{23}^{4})}{9-28C_{24}^{2}C_{23}^{4}-
4C_{23}^{2}+36C_{24}^{2}C_{23}^{2}}~,
\label{eq:rmu24}
\ee
\be
R_{\tau}\simeq \frac{4S_{23}^{2}+28C_{24}^{2}C_{23}^{2}S_{23}^2}
{(23-24C_{23}^2+28C_{23}^4)-4S_{24}^2(9-15C_{23}^2+14C_{23}^4)}~.
\label{eq:rtau24}
\ee

\section{Results}



\begin{figure}[t]
\begin{center}
\includegraphics[width=16.0cm, height=8cm, angle=0]{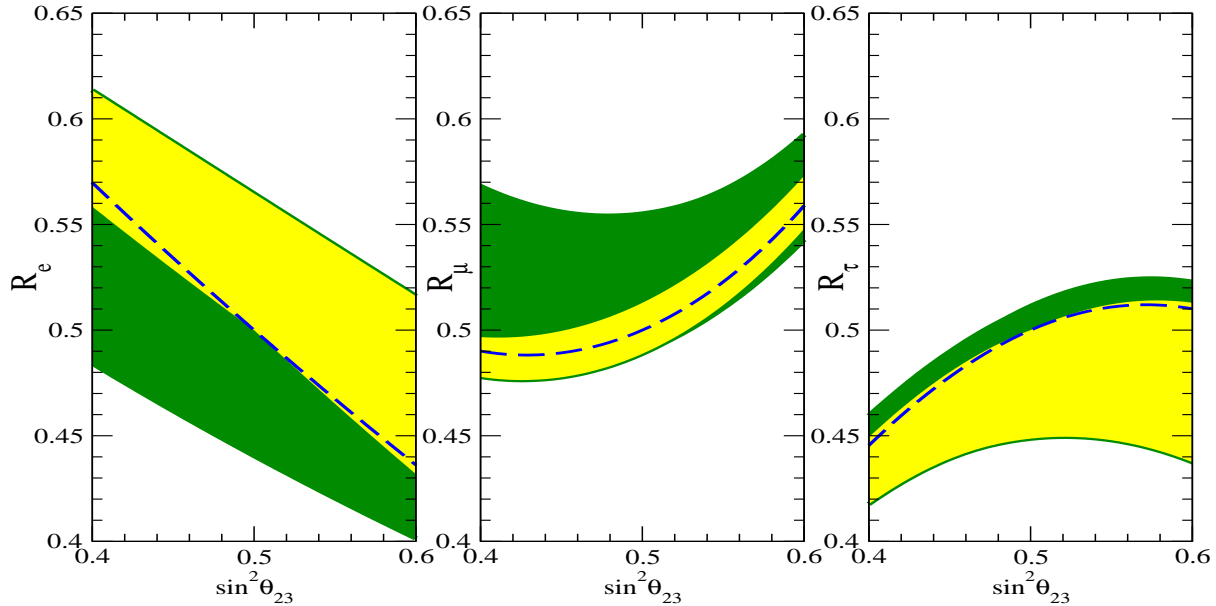}
\caption{\label{fig:TBM23band}
The flux ratios $R_e$, $R_\mu$ and $R_\tau$ versus 
$\sa$, for three active neutrinos. 
The yellow (light) bands corresponds to 
allowed ranges of the flux ratios 
corresponding to all currently $3\sigma$ allowed values 
of $\sss$ and $\sch$ and where $\cos\delta$ is kept fixed 
as 1.
The deep green (dark) bands show the corresponding 
allowed values where $\sss$ and $\sch$ vary in their current 
$3\sigma$ allowed ranges and 
$\cos\delta$ is allowed to vary between $-1$ and 1. 
The upper (lower) boundaries of the yellow and green 
bands almost coincide with each other in the 
left (middle and right) panels.}
\end{center}
\end{figure}

\begin{figure}
\begin{center}
\includegraphics[width=16.0cm, height=7cm, angle=0]{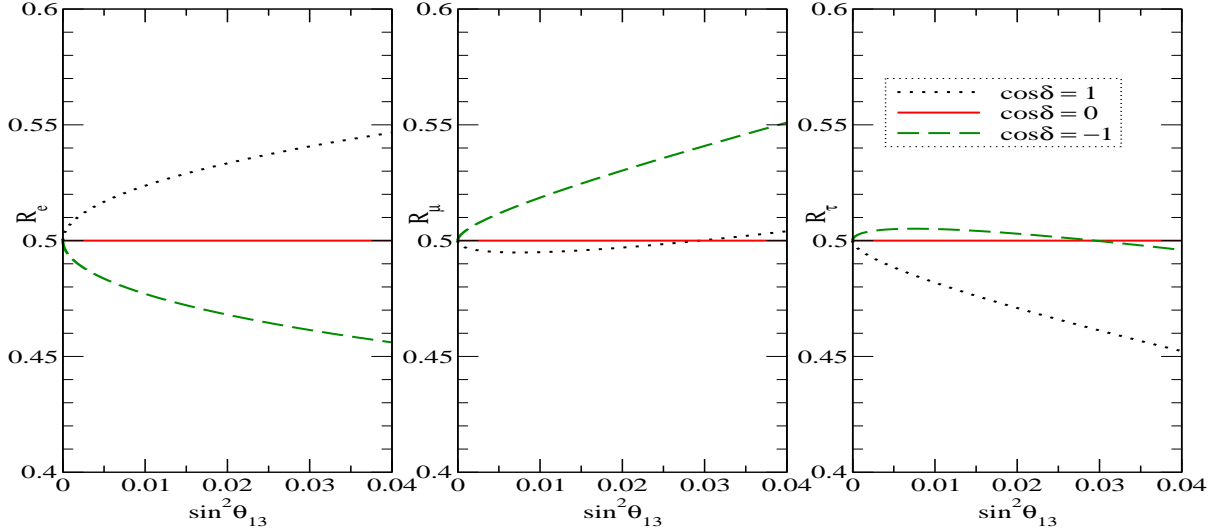}
\caption{\label{fig:TBM13}
The flux ratios $R_e$, $R_\mu$ and $R_\tau$ versus 
$\sch$, for three active neutrinos and when $\sss=1/3$,
$\sa=1/2$. The black dotted lines corresponds to 
$\cos\delta=1$, the red solid lines to $\cos\delta=0$ and 
the green dashed lines to $\cos\delta=-1$.
}
\end{center}
\end{figure}

The mixing angles for both active-active and active-sterile 
mixing are constrained by the results of the neutrino 
oscillation experiments 
\cite{solar,atm,kl,chooz,k2k,minos,lsnd,sbl}. The current 
$3\sigma$ limits on $\theta_{12}$, $\theta_{23}$ and $\theta_{13}$ 
are \cite{limits,limits2}
\be
0.25 < \sss < 0.39~,
\ee
\be
\sin^22\theta_{23} > 0.9~,
\ee
\be
\sch < 0.044~.
\label{eq:sch}
\ee
For the parameterization given by Eq. (\ref{eq:uspara})
which we use for the mixing matrix corresponding to the 
3+1 scenario, 
the mixing angles governing the active-sterile mixing are 
$\theta_{14}$, $\theta_{24}$ and $\theta_{34}$. The mixing angles
$\theta_{14}$ and $\theta_{24}$ are severely constrained by  
LSND \cite{lsnd} and other short baseline data \cite{miniboone,sbl}. 
We refer 
the reader to \cite{maltonischwetz} for the most up-to-date 
analysis of the short baseline experiments. In what follows, 
we will present all our results for 
\be
\sin^2\theta_{14}=0
\ee
and for very small values of $\sin^2\theta_{24}$. Both these mixing 
angles are severely constrained by the current data. 
However, the mixing angle $\theta_{34}$ does not appear in either 
$P_{ee}$ or $P_{e\mu}$ which are probed by most of the short baseline 
experiments like 
Bugey, CDHS, KARMEN, LSND and MiniBooNE. 
This mixing angle appears in the $P_{\mu\tau}$ and $P_{e\tau}$ 
channels. These channels were probed by the 
Chorus and NOMAD experiments at CERN. However, in the 
mass square difference range relevant for LSND,
the constraint on the mixing angle $\theta_{34}$  
are extremely weak. We will therefore allow $\sin^2\theta_{34}$ 
to take all possible values. We reiterate that the 3+1 
mass spectrum for the neutrinos is now comprehensively 
disfavored after the release of the MiniBooNE results. 
Our choice of still using it as an exemplary case stems 
purely from keeping the discussion simple. Our choice of 
taking $\theta_{14}=0$ is also due to the same reason. In fact,
we would have a similar situation  
even if we had taken the 3+2 mass scheme and kept all angles 
except $\theta_{24}$ and $\theta_{34}$ non-zero.

\subsection{Three active neutrinos and deviation from TBM mixing}

In Fig. \ref{fig:TBM23band} we show 
the flux ratios as a function of 
$\sa$ for three active neutrinos. 
The blue dashed lines show the case 
where $\sss$ and 
$\sch$ are kept fixed at their TBM values of $1/3$ and 0
respectively. We see that for this case, $R_e$ varies between 
[0.437-0.569] as $\sa$ changes from [0.4-0.6]. Correspondingly, 
$R_\mu$ and $R_\tau$ vary between [0.488-0.558] 
and [0.446-0.511] respectively. 
We note that $R_e$ depends sharply on $\sa$ and decreases 
linearly as
the value of $\sa$ increases. Both $R_\mu$ and $R_\tau$ 
show a non-linear increase as $\sa$ increases. $R_\mu$ 
increases faster when $\sa > 0.5$, while $R_\tau$ shows a 
sharper dependence when $\sa < 0.5$. 
This means that $R_e$ is a good probe of deviation of 
$\theta_{23}$ from maximality and its octant 
for all values of $\sa$, while 
$R_{\mu}$ is good only for $\sa > 0.5$ and $R_\tau$ for 
$\sa < 0.5$. 
If we allow even $\sss$ and 
$\sch$ to deviate from their TBM mixing values and vary freely 
within their current $3\sigma$ ranges keeping 
$\delta=0$ fixed, then we obtain the 
range of flux ratios shown by the yellow (light) bands 
in Fig. \ref{fig:TBM23band}.
If the CP phase $\delta$ is also allowed to take all possible 
values then the results obtained are shown 
by the green (dark) bands in Fig. \ref{fig:TBM23band}.
We see that the allowed range of $R_e$, $R_\mu$ and $R_\tau$ 
for the currently allowed $3\sigma$ range of 
$\sa$ between [0.34-0.66] 
are respectively 
\be
0.378 \leq &R_e& \leq 0.645 ~,
\nonumber\\
0.476\leq &R_\mu& \leq 0.644 ~,
\nonumber\\
0.379 \leq &R_\tau& \leq 0.524 ~.
\label{eq:3bounds}
\ee
where we allow $\sss$ and $\sch$ to vary within 
their current $3\sigma$ allowed values and $\delta$ over its 
full range. 
If we had three active neutrinos only, since 
we know that  
$\sin^22\theta_{23}>0.9$ at
$3\sigma$, the flux ratios {\it must} be 
restricted within the bounds given by Eq. (\ref{eq:3bounds}).
If we measure a flux ratio outside these ranges in the future 
neutrino telescopes, then that would be a certain indication
of new physics and in particular as we will see in the next subsection,
of presence of extra sterile neutrinos mixed with the active ones. 

In Fig. \ref{fig:TBM13} we show the flux ratios as a function of 
$\sch$, keeping $\sss$ and $\sa$ at their TBM values. 
The black dotted lines in this figure show the 
case for $\cos\delta=1$, the red solid lines  
for $\cos\delta=0$ and green dashed line for $\cos\delta=-1$. 
A comparison of Fig. \ref{fig:TBM13} with 
Fig. \ref{fig:TBM23band} shows that the impact of changing $\sa$ 
from its TBM value of 1/2
is much larger on the flux ratios than 
changing $\sch$ from its TBM value of 0. Therefore, in the 
rest of the paper we only show the effect of 
deviation from TBM mixing by changing $\sa$ from 1/2 and 
keep $\sch$ fixed at 0.

\subsection{Impact of sterile neutrinos}

\begin{figure}
\includegraphics[width=6.0cm, height=7cm, angle=0]{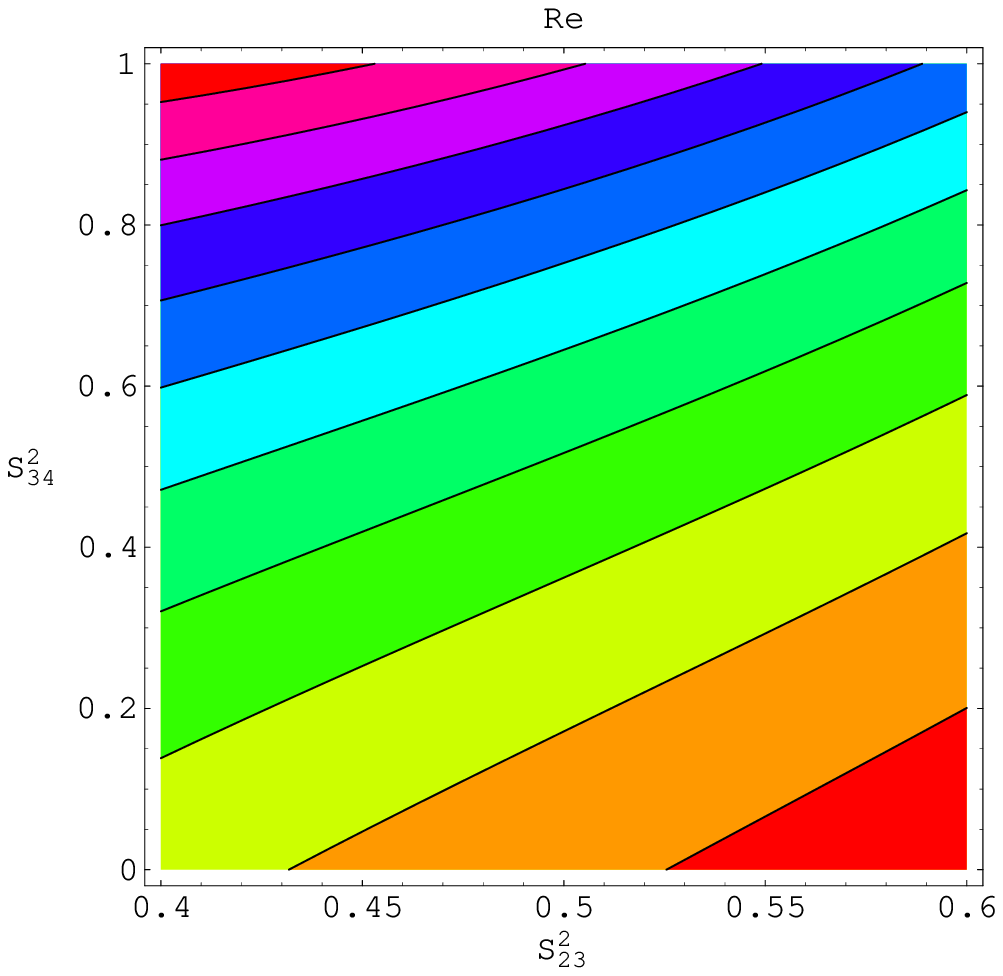}
\vglue -7.0cm,\hglue6.0cm
\includegraphics[width=6.0cm, height=7cm, angle=0]{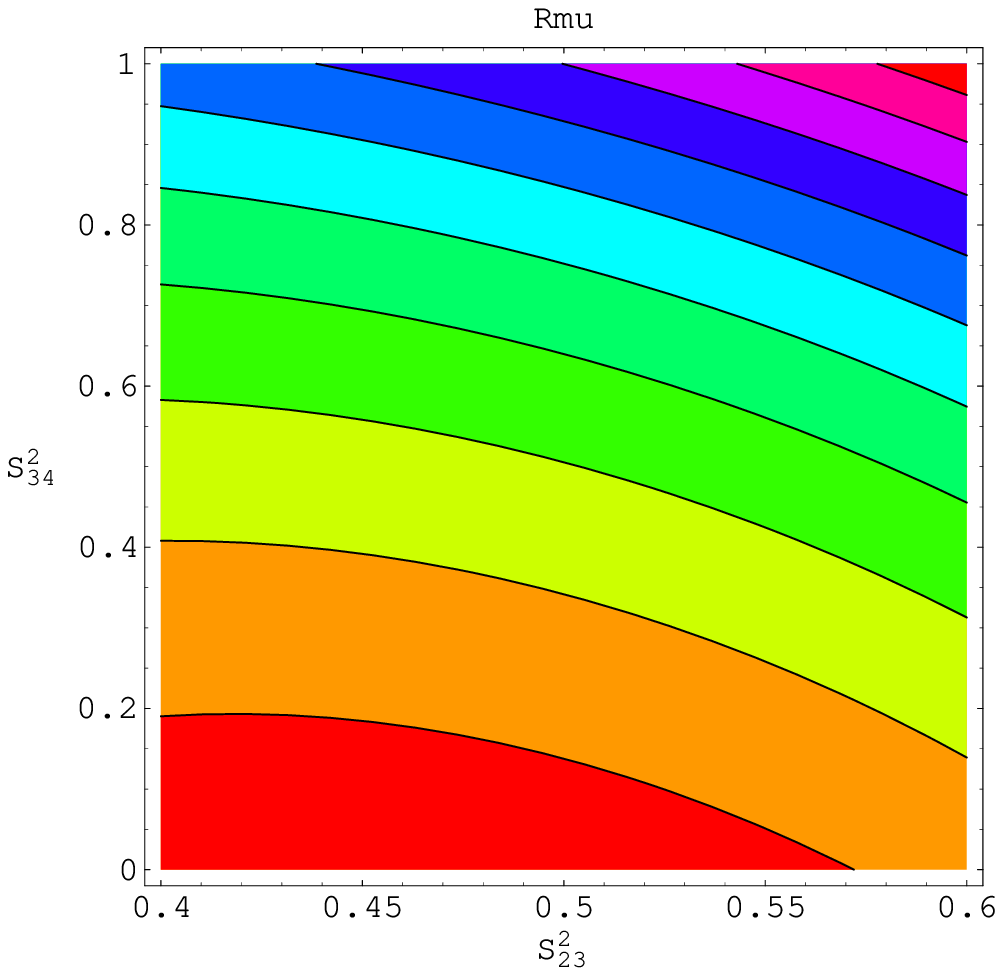}
\vglue -7.0cm,\hglue12.0cm
\includegraphics[width=6.0cm, height=7cm, angle=0]{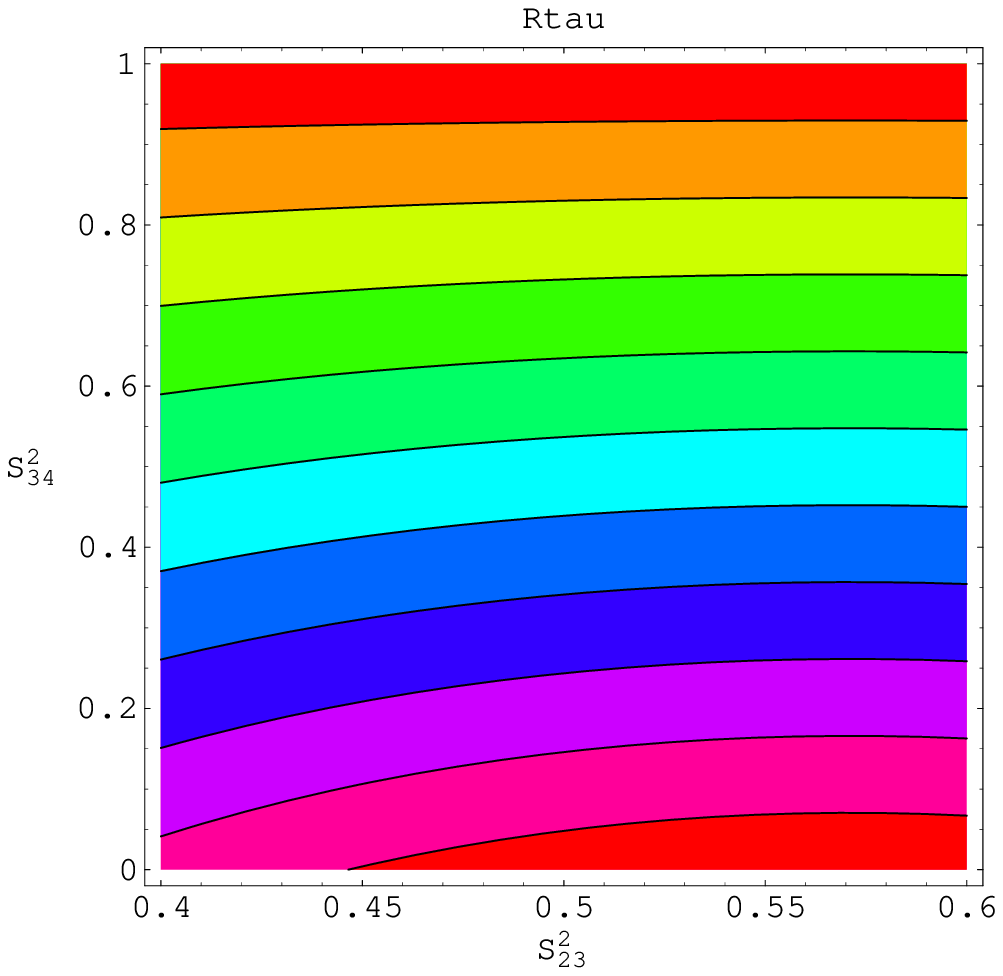}
\caption{\label{fig:contsterile34}
Contour bands of 
the flux ratios $R_e$, $R_\mu$ and $R_\tau$ in the 
$\sa-\sin^2\theta_{34}$ plane. 
The color scheme corresponds to 
flux ratios increasing with
red, orange, yellow, green and so on.
The $R_e$ bands increase from 
0.44 to 1.1 in steps of 0.066, the  $R_\mu$ bands increase from 
0.49 to 1.18 in steps of 0.069, while the  $R_e$ bands increase from 
0.00 to 0.51 in steps of 0.051. 
We have kept $\theta_{14}=0=\theta_{24}$
and all other mixing angles at their TBM mixing values.
}
\end{figure}



\begin{figure}
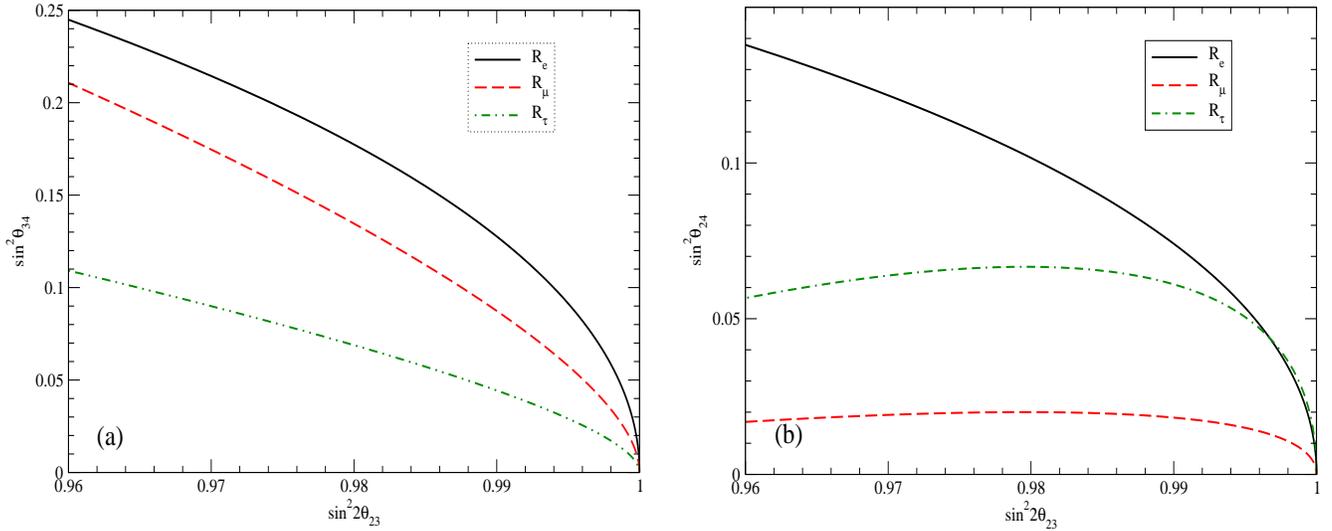

\includegraphics[width=8.5cm, height=7cm, angle=0]{confusion_34_23.eps}
\vglue -7.0cm \hglue 9.cm
\includegraphics[width=8.5cm, height=7cm, angle=0]{confusion_24_23.eps}
\caption{\label{fig:confusion_34_23}
Panel (a) shows the contours 
in the $\sin^22\theta_{23}-\sin^2\theta_{34}$ plane which 
satisfy the 
corresponding condition given by Eq. (\ref{eq:conf_cond}).
Each point on a given line gives the value of 
$\sts$ which along with TBM in active 
part would give the same flux ratio as three active 
neutrinos but with that value for $\sin^22\theta_{23}$.
Panel (b) gives the corresponding contours 
in the $\sin^22\theta_{23}-\sin^2\theta_{24}$ plane, showing 
the condition Eq. (\ref{eq:conf_cond24}).
}
\end{figure}

For the sake of illustration we begin by presenting results 
for the case where the active-sterile 
mixing angles $\sin^2\theta_{14}=0$ and 
$\sin^2\theta_{24}=0$. For the active-active mixing 
angles we take $\sss$ and $\sch$ at their TBM values of 1/3 and 
0 respectively and allow only $\sa$ to vary on both sides of 
its TBM value of 1/2. In other words, deviation
from TBM mixing is caused 
by a non-maximal $\theta_{23}$, while presence of 
sterile neutrino and its impact comes through the non-zero 
mixing angle $\theta_{34}$.  

In Fig. \ref{fig:contsterile34} we show the contour bands of 
the flux ratios $R_e$, $R_\mu$ and $R_\tau$ in the 
$\sa$-$\sin^2\theta_{34}$ plane. 
The color scheme corresponds to 
flux ratios increasing with
red, orange, yellow, green and so on.
The $R_e$ bands increase from 
0.44 (lower right-hand corner)
to 1.1 (upper left-hand corner) 
in steps of 0.066, the  $R_\mu$ bands increase from 
0.49 (lower left-hand corner)
to 1.18 (upper right-hand corner)
in steps of 0.069, while the  $R_\tau$ bands increase from 
0.00 (top)
to 0.51 (bottom) in steps of 0.051. 
As noted in the previous subsection, 
we see from the figure that $R_e$ has a very 
sharp dependence on $\sa$ and its value decreases as  
$\sa$ increases. 
$R_\mu$ shows a milder dependence and its 
value increases with $\sa$, while $R_\tau$ is almost 
independent of the value of $\sa$ for $\sa > 0.5$ and 
shows an extremely  mild dependence for $\sa<0.5$. 
All flux ratios depend 
strongly on the value of $\sin^2\theta_{34}$. The flux ratios 
$R_e$ and $R_\mu$ increase as $\sin^2\theta_{34}$ 
increases, while $R_\tau$ decreases. 
This can be seen also from 
Eqs. (\ref{eq:re_24zero}), (\ref{eq:rmu_24zero}) and 
(\ref{eq:rtau_24zero}), where $\sin^2\theta_{34}$ 
decreases the denominators of $R_e$ and $R_\mu$, 
while it decreases the numerator of $R_\tau$.
The figure also 
shows that the dependence of $R_\tau$ on $\sin^2\theta_{34}$
is linear, while that
of $R_e$ and $R_\mu$ are non-linear. This 
is corroborated by the  
Eqs. (\ref{eq:rtau_24zero}), 
(\ref{eq:re_24zero}) and (\ref{eq:rmu_24zero}) respectively. 
%
The most important thing we can see from this figure 
is the following: 
the same value of 
$R_e$, $R_\mu$ and $R_\tau$ can be obtained for different set of 
values of $\sa$ and $\sts$. This gives us degenerate solutions 
in the $\sa$-$\sts$ plane. 
This will therefore lead to confusion\footnote{This is just one 
example in which we show this confusion. This confusion is expected 
even in the most general case where there might be 
more sterile neutrinos, with all mixing angles and CP phases 
non-zero.}
between 
the case where there are only three active neutrinos with 
deviation from TBM mixing and the case where TBM holds in the three 
active regime but there are extra sterile neutrinos which are 
mixed with the active neutrinos.  
This degeneracy is obviously more 
pronounced when the correlation between $\sa$ and $\sts$ is 
greater. Therefore, the largest effect is expected in $R_e$ and 
smallest in $R_\tau$. 

To further illustrate the confusion between 
deviation from TBM mixing and 
presence of sterile neutrinos we present
Fig. \ref{fig:confusion_34_23}(a), which shows 
in the $\sin^22\theta_{23}$-$\sts$ plane
the contours for the condition
\be
R_\alpha(\sa\neq 0.5,\sts=0) = R_\alpha(\sa=0.5,\sts\neq 0)~,
\label{eq:conf_cond}
\ee
where $\alpha=e$, $\mu$ or $\tau$. The L.H.S of Eq. 
(\ref{eq:conf_cond}) gives the flux ratios predicted by
deviation from TBM mixing for three active neutrinos only, 
while the R.H.S corresponds 
to TBM mixing in the active part but with extra sterile 
neutrinos. We remind the reader that $\sss=1/3$, $\sch=0$,
$\sin^2\theta_{14}=0$ and $\sin^2\theta_{24}=0$ on 
both sides of  Eq. (\ref{eq:conf_cond}).
Using Eq. (\ref{eq:re_24zero})-(\ref{eq:rtau_24zero})
condition (\ref{eq:conf_cond}) leads to
\be
S_{34}^{2}&=&\frac{12-24S_{23}^{2}}{13-8S_{23}^{2}} 
\label{eq:cond_conf1}
\ee
for $\alpha=e$, 
\be
S_{34}^{2}&=&\frac{15-72S_{23}^{2}+84S_{23}^{4}}
{14-24S_{23}^{2}+28S_{23}^{4}} 
\label{eq:cond_conf2}
\ee
for $\alpha=\mu$ and 
\be
S_{34}^{2}&=&\frac{27-96S_{23}^{2}+84S_{23}^{4}}
{27-32S_{23}^{2}+28S_{23}^{4}} 
\label{eq:cond_conf3}
\ee
for $\alpha=\tau$.
We have plotted Eqs. (\ref{eq:cond_conf1})-(\ref{eq:cond_conf3}) in 
Fig. \ref{fig:confusion_34_23}(a), where the solid line 
corresponds to the condition
for $R_e$,
dashed line for $R_\mu$ and dot-dashed line for $R_\tau$.
Any point on a given line in this figure gives the value of 
$\sa$ and $\sts$ such that Eq. (\ref{eq:conf_cond}) is 
satisfied. For instance, the case with three active neutrinos 
only and with $\sin^22\theta_{23}=0.96$ predicts a value of $R_e$ 
which can be obtained with TBM mixing 
in the active sector plus an extra sterile neutrino 
with $\sts=0.245$. We note that 
for the same $\sin^22\theta_{23}$, the values of $\sts$ 
for the sterile case which can also simultaneously satisfy the 
condition for $R_\mu$ and $R_\tau$ are different, the 
corresponding values of $\sts$ being  
$0.21$ and $0.11$ respectively. This means that 
in principle if we 
knew what is the true value of $\sts$, we would be able to distinguish 
the results for the sterile case with that for active neutrinos only 
with deviation from TBM mixing, if we measured all the three flux 
ratios simultaneously. 
However, in practice we do not have 
any precise information on this mixing angle. Therefore, this 
confusion will be hard to solve even if we had measurement on 
all the three flux ratios at neutrino telescopes. 

The second very important aspect evident from Figs. 
\ref{fig:contsterile34} 
is that 
the values of the flux ratios can easily exceed the 
range given in Eq. (\ref{eq:3bounds}). 
For instance, we see 
that values of $R_e>0.645$ 
is possible for $\sa>0.4$ and $\sts>0.25$. Note that 
we have kept $\sch=0$, $\sss=1/3$ and all CP phases to 0. 
In fact for this case, when 
$\sts=0$ the range of $R_e$ corresponds to [0.437-0.569],
as in Fig.  \ref{fig:TBM23band} for $\delta=0$. 
We can draw similar conclusions from 
$R_\mu$ and $R_\tau$ using the  
middle and right panels of Fig. \ref{fig:contsterile34}. 
If the measured flux ratios correspond to such extreme values,
then this could be a signature for extra sterile neutrinos. 

\begin{figure}[t]
\includegraphics[width=6.0cm, height=7cm, angle=0]{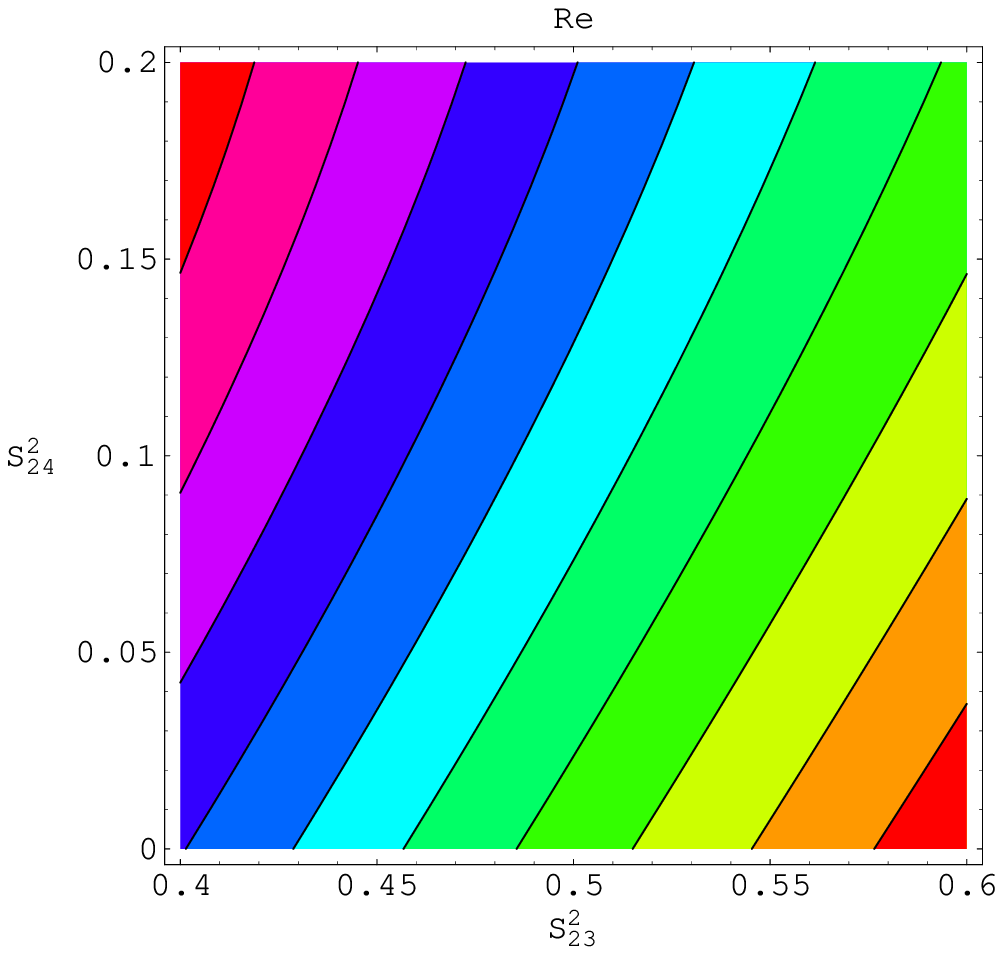}
\vglue -7.0cm,\hglue6.0cm
\includegraphics[width=6.0cm, height=7cm, angle=0]{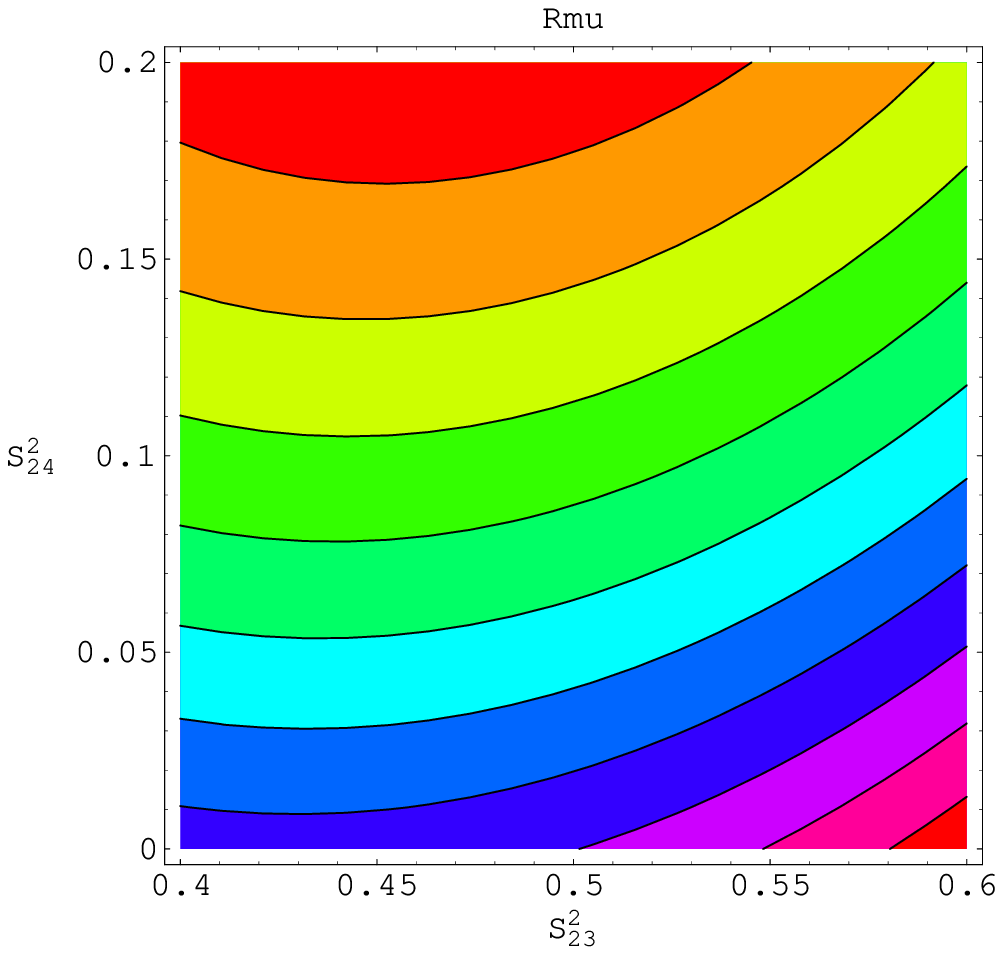}
\vglue -7.0cm,\hglue12.0cm
\includegraphics[width=6.0cm, height=7cm, angle=0]{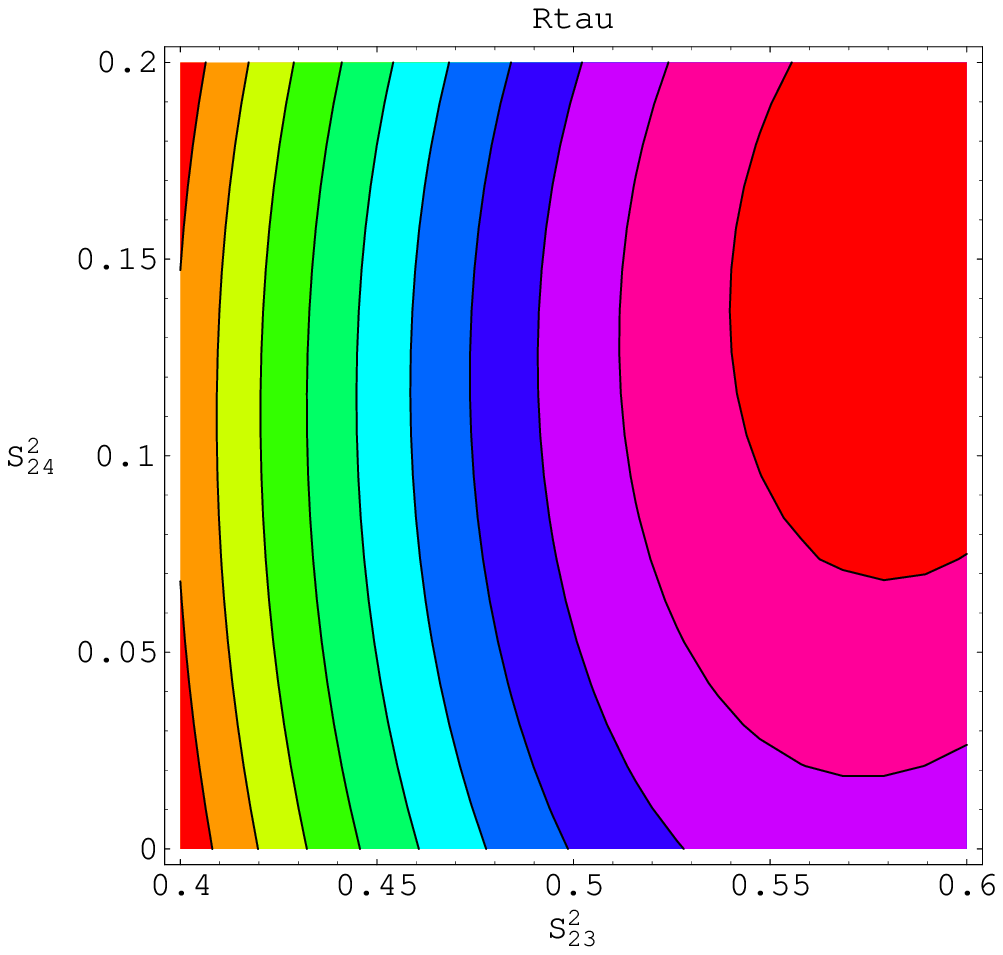}
\caption{\label{fig:contsterile24}
Contour bands of 
the flux ratios $R_e$, $R_\mu$ and $R_\tau$ in the 
$\sa-\sin^2\theta_{24}$ plane. 
The color scheme corresponds to 
flux ratios increasing with
red, orange, yellow, green and so on.
The $R_e$ bands increase from 
0.436 
to 0.610 
in steps of 0.017, the  $R_\mu$ bands increase from 
0.447 
to 0.559 
in steps of 0.011, while the  $R_\tau$ bands increase from 
0.445 
to 0.526 
in steps of 0.008. 
We have kept $\theta_{14}=0=\theta_{34}$
and all other mixing angles at their TBM mixing values.
}
\end{figure}
%
\begin{figure}[t]
\includegraphics[width=5.5cm, height=4.99cm, angle=0]{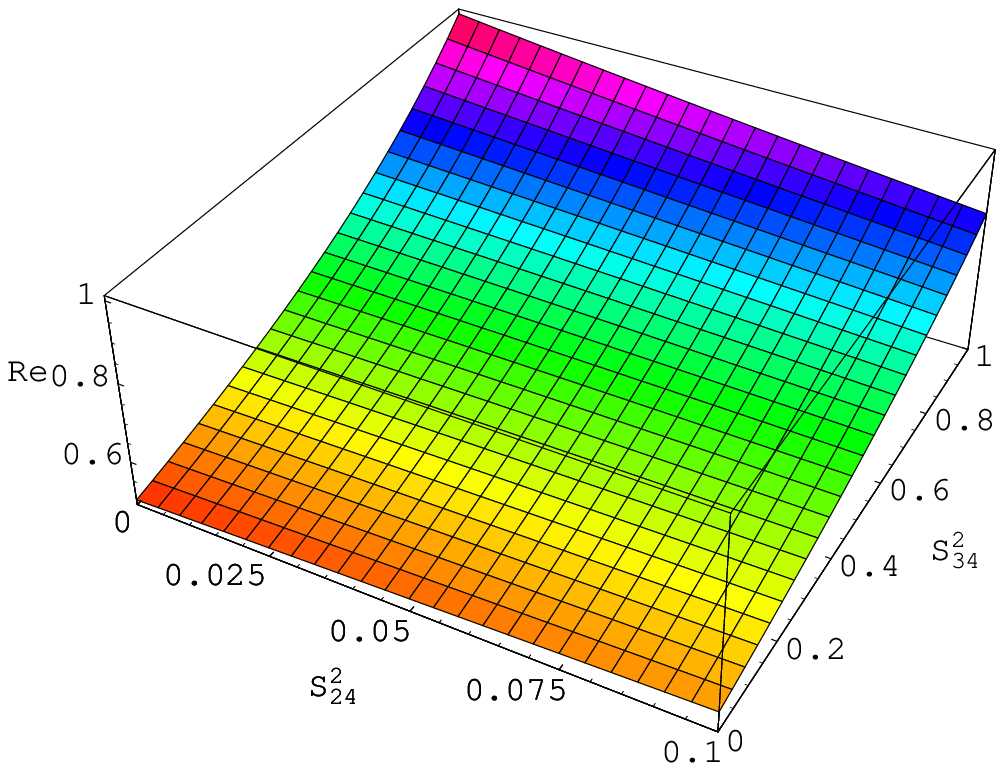}
\vglue -4.99cm,\hglue5.7cm
\includegraphics[width=5.5cm, height=4.99cm, angle=0]{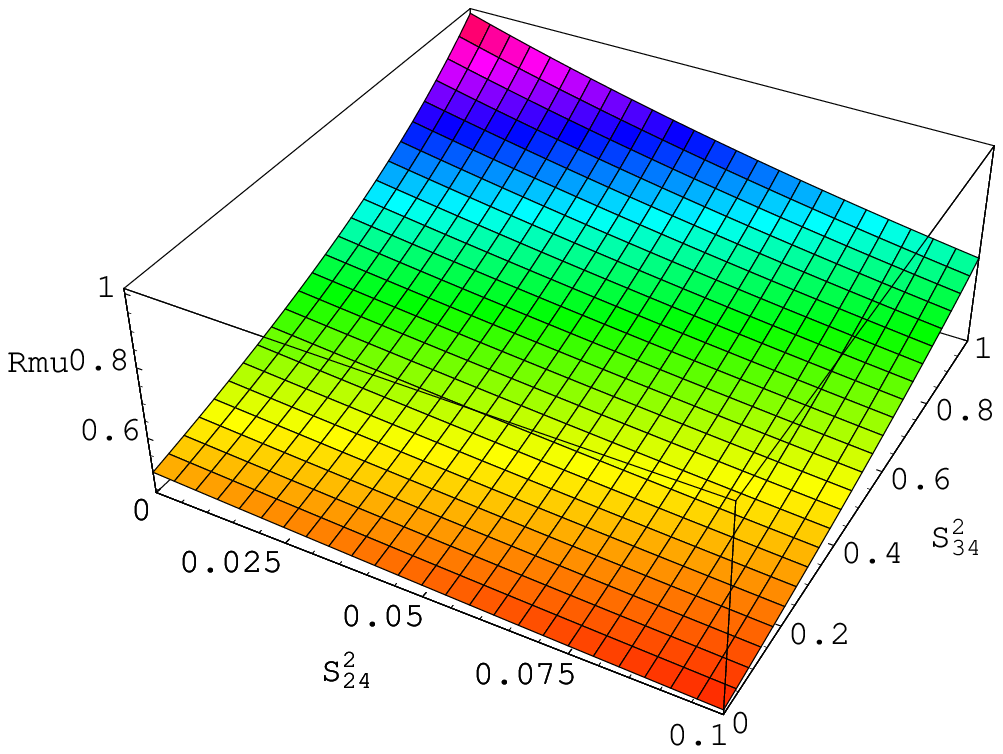}
\vglue -4.99cm,\hglue11.4cm
\includegraphics[width=5.7cm, height=4.99cm, angle=0]{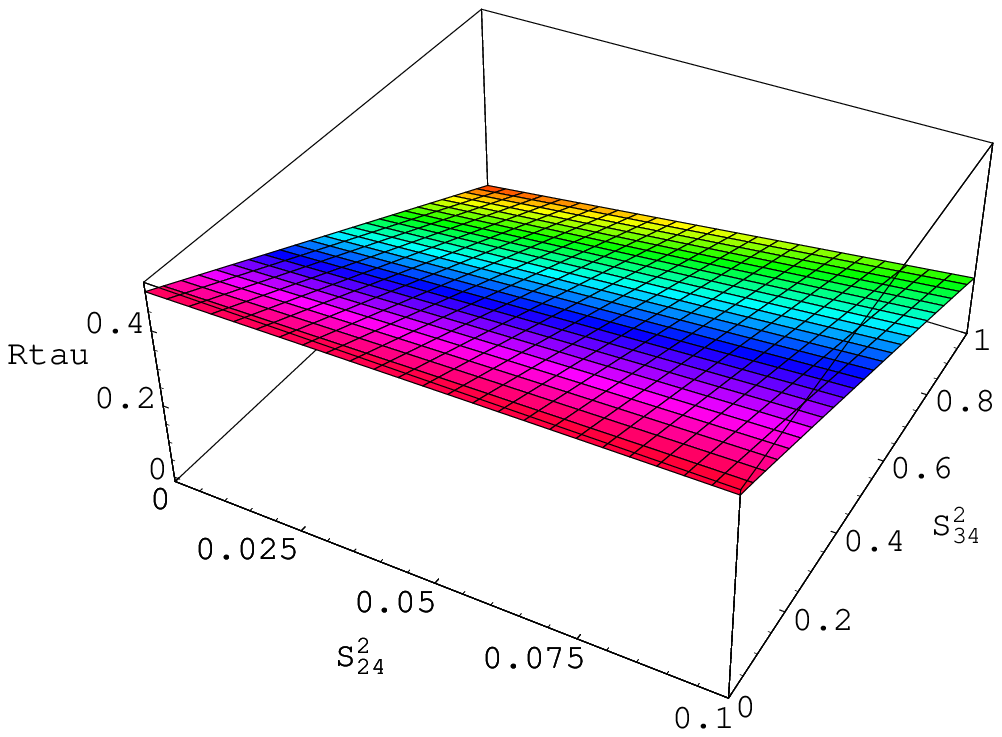}
\caption{\label{fig:R_24_34_fixed23}
Three dimensional plots showing the flux ratios 
$R_e$ (left panel), $R_\mu$ (middle panel) and 
$R_\tau$ (right panel), as a function of 
$\sin^2\theta_{24}$ and $\sts$. 
The angles $\theta_{13}=0=\theta_{14}$, $\sss=1/3$ and $\sa=0.5$.
}
\end{figure}

In Fig. \ref{fig:contsterile24} we show the contours of constant 
flux ratios in the $\sa$-$\sin^2\theta_{24}$ plane, keeping
$\theta_{34}=0$, $\theta_{14}=0$, $\theta_{13}=0$ and 
$\sss=1/3$. The color scheme is same as in Fig. 
\ref{fig:contsterile34}. 
The $R_e$ bands increase from 
0.436 (lower right-hand corner)
to 0.610 (upper left-hand corner) 
in steps of 0.017, the  $R_\mu$ bands increase from 
0.447 (upper left-hand side)
to 0.559 (lower right-hand corner)
in steps of 0.011, while the  $R_\tau$ bands increase from 
0.445 (left-hand side)
to 0.526 (right-hand side) in steps of 0.008. 
Note that even though we have shown results up to 
$\sin^2\theta_{24}=0.2$, this mixing angle is severely constrained
and  
is expected to be $\ltap 0.05$. We note 
from the figure 
that $R_e$ increases with $\sin^2\theta_{24}$, as the 
denominator in Eq. (\ref{eq:re24}) decreases much faster
with $\sin^2\theta_{24}$ than the numerator. 
On the 
other hand $R_\mu$ 
decreases with $\sin^2\theta_{24}$
because the numerator in Eq. (\ref{eq:rmu24}) 
decreases with $\sin^2\theta_{24}$ faster than the 
denominator. For $R_\tau$ the dependence is more 
complicated. The Fig. \ref{fig:confusion_34_23}(b)
illustrates the confusion that could arise between a 
three-generation scheme with deviation from TBM mixing 
with the one with sterile neutrinos. In the panel (b) 
of this figure we show the contours corresponding to the
condition
\be
R_\alpha(\sa\neq 0.5,\sin^2\theta_{24}=0) = 
R_\alpha(\sa=0.5,\sin^2\theta_{24}\neq 0)~,
\label{eq:conf_cond24}
\ee
where $\alpha=e$, $\mu$ or $\tau$. As before,
the L.H.S of Eq. 
(\ref{eq:conf_cond24}) gives the flux ratios predicted by
deviation from TBM mixing for three active neutrinos only, 
while the R.H.S corresponds 
to TBM mixing in the active part but with extra sterile 
neutrinos. We reiterate that $\sss=1/3$, $\sch=0$,
$\sin^2\theta_{14}=0$ and $\sin^2\theta_{34}=0$ on 
both sides of  Eq. (\ref{eq:conf_cond24}). From Eqs.
(\ref{eq:re24}), (\ref{eq:rmu24}) and (\ref{eq:rtau24}) we 
get the conditions 
\begin{eqnarray}
S_{24}^{2}=\frac{4-8S_{23}^{2}}{9-8S_{23}^{2}}
\end{eqnarray} 
for $\alpha=e$, 
\begin{eqnarray}
S_{24}^{2}=\frac{5-24S_{23}^{2}+28S_{23}^{4}}{28S_{23}^{4}-24S_{23}^{2}-2}
\end{eqnarray}
for $\alpha=\mu$ and 
\begin{eqnarray}
S_{24}^{2}=\frac{9-32S_{23}^{2}+28S_{23}^{4}}{7-32S_{23}^{2}+28S_{23}^{4}}
\end{eqnarray} 
for $\alpha=\tau$. These conditions are plotted in 
Fig. \ref{fig:confusion_34_23}(b).

In the general scenario of course we expect both 
$\sin^2\theta_{24}$ and $\sts$ to be non-zero. However, we 
still keep $\sin^2\theta_{14}$ to be zero for simplicity. 
In Fig. \ref{fig:R_24_34_fixed23} we present three dimensional 
plots showing the flux ratios as a function of 
$\sin^2\theta_{24}$ and $\sts$ for 
$\sa=0$. The left panel shows $R_e$, 
middle panel shows $R_\mu$ and right panel 
shows $R_\tau$.
We have kept $\theta_{13}=0$ and $\sss=1/3$. 

\section{Discussions}

\begin{table}[t]
\begin{center}
\begin{tabular}{|c|c|c|c|} \hline
Model & $\sin^2\theta_{24}$ & $\sin^2\theta_{34}$ & Flavor at Earth \cr
\hline
Standard & -- & -- & $1:1:1$ \cr
Decay (NH) & -- & -- & $4:1:1$ \cr
Decay (IH) & -- & -- & $0:1:1$ \cr
CPT [$\frac{1}{2}(\Phi_{std} + \bar\Phi_{CPTV})$] & -- & -- & 
$1.07:1:0.79$ \cr
CPT + Decay (NH) [$\frac{1}{2}(\bar\Phi_{CPTV})$ only] & -- & -- & $2.43:1:0.14$ \cr
CPT + Decay (IH) [$\frac{1}{2}(\bar\Phi_{CPTV})$ only] & -- & -- &
$0.04:1:3.30$ \cr
Pseudo-Dirac (only 1) & -- & -- & $0.73:1:1$ \cr
Pseudo-Dirac (only 2) & -- & -- & $1:1:1$ \cr
Pseudo-Dirac (2 and 3) & -- & -- & $1.43:1:1$ \cr
Sterile & 0.04 & 0.0 & $1.05:1:1.04$ \cr
Sterile & 0.04 & 1.0 & $1.05:1:0.12$ \cr
Sterile & 0.1 & 0.0 & $1.12:1:1.08$ \cr
Sterile & 0.1 & 1.0 & $1.12:1:0.32$ \cr
\hline
\end{tabular}
\caption{\label{tab:compare}
Comparison of the flavor ratios predicted by the different 
new physics scenarios considered in the literature with the 
standard three flavor picture (first row) and 
presence of extra sterile neutrinos (bottom rows). 
Note that for the CPTV+decay models, the flavor ratios correspond 
to antineutrinos only. For pseudo-Dirac neutrinos we show only
two cases, one where only the oscillations due to the 
second mass splitting is averaged out and another where 
oscillations due to both second and third splitting are
averaged out. We refer the reader to \cite{uhepseudo} for 
the complete list. Note that we have used mixing angles 
corresponding for TBM mixing for the ratios 
given in this table. This is why they might differ slightly 
from the ones given in \cite{uhedecay,uhecpt,uhepseudo}.
}
\end{center}
\end{table}

In this paper we have considered presence of sterile 
neutrinos as the only signature of physics beyond the 
standard paradigm.
Other forms of new physics could also manifest themselves in 
the observed flux ratios of the ultra high energy neutrinos.
In particular, neutrino decay predicts spectacular 
results for the ratios, where one expects the flavor ratio $4:1:1$ 
for the normal hierarchy\footnote{The authors of 
\cite{uhedecay} get $6:1:1$ because they take 
$\sss=1/4$. The ratio given above is obtained 
for TBM mixing for which $\sss=1/3$.}
and $0:1:1$ for the inverted 
hierarchy \cite{uhedecay}. Authors of \cite{uhecpt}
considered the effect of  
CPT violation on the flavor ratios, 
for both stable as well as 
unstable neutrinos. In \cite{uhepseudo} 
prospects of probing the existence of pseudo-Dirac
neutrinos was discussed. In Table \ref{tab:compare}
we show the 
predicted flavor ratios for all these new physics 
scenarios and compare them with that expected for
sterile neutrinos. 
We note that the largest deviation for $\phi_e/\phi_\mu$
comes for the neutrino decay model, both with normal (NH) 
and inverted (IH) 
hierarchy. The ratio $\phi_\mu/\phi_\tau$ however for this 
model stays at 1, just like in the standard neutrino case.
This ratio is predicted to be very different for the case where 
we have CPT violation (with and without neutrino decay) and 
where we have sterile neutrinos with large $\sin^2\theta_{34}$.
For sterile neutrinos, this ratio decreases almost linearly with 
$\sin^2\theta_{34}$, the slope being  
determined by $\sin^2\theta_{24}$. 
We can see that for very large $\sin^2\theta_{34}$, 
the $\phi_\mu/\phi_\tau$ ratio predicted by sterile neutrinos is
very similar to that expected for antineutrinos with CPT violation 
plus decay 
and NH. In fact, since  $\phi_\mu/\phi_\tau$ for sterile neutrinos starts 
from a large value and decreases to a small number.
Therefore, for some 
range of values of $\sin^2\theta_{34}$ one would get a 
$\phi_\mu/\phi_\tau$
similar to the one predicted by CPT violation with 
stable neutrinos as well. Therefore, we see that there is ample 
scope of confusing sterile neutrino signatures in ultra high energy
flavor ratio with that of CPT violation. We also note that the 
$\phi_e/\phi_\mu$ ratio predicted by sterile neutrinos 
could be close to some of the cases for pseudo-Dirac neutrinos.

So far we have presented all results for
ratios defined in terms 
of the ultra high energy neutrino fluxes arriving at Earth. 
However, one should bear in mind that what is physically 
relevant as far as the measurement is concerned, are 
the ratio of the corresponding number of events in the 
neutrino telescope. 
The number of events expected in the 
$\numu$, $\nue$ and $\nutau$ channels for an assumed 
normalization for the ultra high energy neutrino flux 
was performed in \cite{uheflavor}. 
While $\numu$ are easiest to detect as 
the resultant muons leave distinct tracks in the detector,
the $\nue$ are detected by observing the electromagnetic showers 
the electrons create. 
The efficiency of detecting electron 
events is generally lower than that for muons and it decreases 
as the neutrino energy increases. The threshold for muon events 
is expected to be about $10^2$ GeV while that for electrons would 
be about $10^3$ GeV. 
The $\nutau$ are detectable through the so-called double-bang 
and lollipop events \cite{doublebang}
only at relatively higher energies of about $10^6$ GeV. 
Their detection efficiency would become comparable 
to that for electrons for energies larger than  $10^7$ GeV
\cite{uheflavor}.
In all our results we had tacitly given 
equal weightage to all three flavors. However from the 
discussion above, it is clear that the behavior of the 
flavor ratios would change, once we used the event ratios 
rather than the flux ratios, at least for neutrino energies 
$<< 10^7$ GeV. In fact, for neutrino energies 
below $10^6$ GeV, we should not include 
the $\nutau$ flux at all and work with just the 
$\phi_e/\phi_\mu$ ratio. Above energies of $10^7$ GeV 
we could in principle work with all three flavors, 
however the neutrino flux falls 
steeply with energy and we expect only a handful of 
events at these very high energies.
We have seen in Table \ref{tab:compare} that 
sterile components cause 
largest 
deviation from the standard model prediction for 
$\phi_\tau$. To see this we would necessarily 
need observations at higher energies, where statistical 
uncertainty could be the biggest challenge. 
A detector 
much larger than IceCube would be 
probably required for getting an unambiguous 
signature of sterile species in ultra high energy neutrinos. 
This constraint is applicable to most cases which rely 
on flavor ratios involving the $\nutau$.
However, we reiterate that even though 
unambiguous {\it signal} for sterile neutrinos in a 
km$^3$ detector such as IceCube might prove to be 
difficult, sterile neutrinos will definitely 
cause confusion with the 
signal for deviation from TBM mixing, 
and this conclusion is valid for all 
neutrino telescopes.

\section{Summary and Conclusions}

Neutrino telescopes offer a possible way of constraining the 
mixing angles. In particular, considerable interest has been 
generated of late in using the flux/flavor ratios of 
ultra high energy neutrinos in probing the deviation of the 
mixing matrix from the TBM mixing ansatz. Since it 
has been observed that the effect of deviating $\sss$ from 
1/3 on the flux ratios is not large, we kept $\sss$ 
fixed at its TBM value of $1/3$. Deviation from TBM mixing 
can then be caused either by $\sch \neq 0$ or $\sa \neq 1/2$. 
We showed that the effect of changing $\sch$ from zero 
is small on the flux ratios. We argued that the maximum impact 
on the flux ratios arise when we change
the mixing angle $\sa$ from 1/2. Therefore, we 
presented most results by varying the value of $\sa$ on 
both sides of its TBM value.

We presented the expressions for the flux ratios assuming 
that there was one extra sterile neutrino and the mass spectrum 
followed a 3+1 pattern. Even though we are aware that this mass 
spectrum is now disfavored following the recently declared 
MiniBooNE results and the mass spectrum allowed is the 
3+2 scheme, we used the 3+1 scheme as an exemplary case for 
simplicity. Our formalism can be easily extended to the 3+2 
scenario, only then one has to contend with many more 
parameters. We showed that even in a very simplified picture 
where only one active-sterile mixing angle $\sts$ 
(or $\sin^2\theta_{24}$) was kept 
non-zero, it was possible to easily confuse the predicted 
flux ratios for the deviation from TBM case with three active 
neutrinos with the case where TBM holds in the three active sector 
but we have one extra sterile neutrino. We also showed that 
for very large values of the mixing angle $\sts$, which as of now is 
almost unconstrained, we get predictions for the flux 
ratios that are completely out of the possible range 
predicted by three active neutrinos. If the measured 
flux ratio really conformed with such extreme values, 
this would then be a smoking gun signal for sterile neutrinos 
which are heavily mixed with active neutrinos. We also 
presented results where we took $\sin^2\theta_{24}$ to be the 
only non-zero active-sterile mixing angle. Finally, we 
showed results for two non-zero active-sterile mixing angles,
$\sts$ and  $\sin^2\theta_{24}$. 

We stress that our results would be valid 
even if a sterile neutrino is not needed to explain the LSND 
data. Since for ultra high energy neutrinos coming from astrophysical 
sources, the oscillatory terms anyway average out to zero 
irrespective of the value of $\Delta m^2$, a sterile neutrino 
with any value for $\Delta m^2$ would give such signatures 
in the neutrino telescopes if the  mixing angle such as $\theta_{34}$, 
which has not yet been checked in terrestrial experiments,  
was non-zero and large.

In conclusion, if the values of the measured flux ratios in 
the future neutrino telescopes turns out be different from 
their TBM prediction of 1/2, it would still not be a foolproof
evidence for deviation from TBM mixing. Sterile neutrinos with 
reasonably small mixing with active neutrinos could mimic a 
similar response, even when the active sector conforms to TBM 
mixing. 
One will never be able to tell one scenario from 
the other by measuring the flux ratios of ultra high energy
neutrinos,
if the measured flux ratios were within the range 
predicted by the standard picture with three active neutrinos.
However, if the 
measured ratios turn out to be clearly outside the 
range predicted by the three active case, we would have a 
signal for the existence of sterile neutrinos. One 
probably needs a larger detector with sufficient 
statistics for $\nutau$ events for an unambiguous signal.



\end{document}